\documentclass[russian,english]{extarticle}
\usepackage[T2A,T1]{fontenc}
\usepackage[koi8-r,latin9]{inputenc}
\usepackage{amsmath}
\usepackage{amssymb}
\usepackage{esint}

\makeatletter

\DeclareRobustCommand{\cyrtext}{%
  \fontencoding{T2A}\selectfont\def\encodingdefault{T2A}}
\DeclareRobustCommand{\textcyr}[1]{\leavevmode{\cyrtext #1}}
\AtBeginDocument{\DeclareFontEncoding{T2A}{}{}}

\newcommand{\lyxmathsym}[1]{\ifmmode\begingroup\def\b@ld{bold}
  \text{\ifx\math@version\b@ld\bfseries\fi#1}\endgroup\else#1\fi}

\@ifundefined{date}{}{\date{}}
\AtBeginDocument{\DeclareFontEncoding{T2A}{}{}}\AtBeginDocument{\DeclareFontEncoding{T2A}{}{}}\AtBeginDocument{\DeclareFontEncoding{T2A}{}{}}\AtBeginDocument{\DeclareFontEncoding{T2A}{}{}}\AtBeginDocument{\DeclareFontEncoding{T2A}{}{}}\AtBeginDocument{\DeclareFontEncoding{T2A}{}{}}

\newtheorem{theorem}{Theorem}
\newtheorem{proposition}{}

\newtheorem{lemma}{Lemma}

\usepackage{babel}

\makeatother

\usepackage{babel}
\begin{document}

\title{Regular continuum systems of point particles. I: systems without
interaction}

\author{Chubarikov V. N., Lykov A. A., Malyshev V. A.\thanks{Lomonosov Moscow State University, Faculty of Mechanics and Mathematics,
Vorobyevy Gory 1, Moscow, 119991, Russia} }
\maketitle
\begin{abstract}
Normally, in mathematics and physics, only point particle systems,
which are either finite or countable, are studied. We introduce new
formal mathematical object called regular continuum system of point
particles (with continuum number of particles). Initially, each particle
is characterized by the pair: (initial coordinate, initial velocity)
in $R^{2d}$. Moreover, all initial coordinates are different and
fill up some domain in $R^{d}$. Each particle moves via normal Newtonian
dynamics under influence of some external force, but there is no interaction
between particles. If the external force is bounded then trajectories
of any two particles in the phase space do not intersect. More exactly,
at any time moment any two particles have either different coordinates
or different velocities. The system is called regular if there are
no particle collisions in the coordinate space. 

The regularity condition is necessary for the velocity of the particle,
situated at a given time at a given space point, were uniquely defined.
Then the classical Euler equation for the field of velocities has
rigorous meaning. Though the continuum of particles is in fact a continuum
medium, the crucial notion of regularity was not studied in mathematical
literature.

It appeared that the seeming simplicity of the object (absence of
interaction) is delusive. Even for simple external forces we could
not find simple necessary and sufficient regularity conditions. However,
we found a rich list of examples, one-dimensional and mufti-dimensional,
where we could get regularity conditions on different time intervals.
In conclusion we formulate many unsolved problems for regular systems
with interaction.
\end{abstract}
.


\section{Introduction}

Now we give the exact definition of the central object we will study.

Regular continuum system $\mathbf{M}_{T}$ of point particles is the
set of subsets $\Lambda_{t}\in R^{d}$ enumerated by the time moments
$t\in[0,T),0<T\leq\infty$. Moreover, $\Lambda_{0}$ is assumed to
be the closure of some open connected subset of $R^{d}$ with piece-wise
smooth boundary $\partial\Lambda_{0}$. Each point of this domain
is considered as a <<material particle>> of infinitely small mass.
The dynamics is defined by the system of one-to-one mappings (diffeomorphisms)
$U_{t}=U_{0,t}:\Lambda_{0}\rightarrow\Lambda_{t},t\in[0,\lyxmathsym{\textcyr{\char210}})$.
All these mappings are assumed to be sufficiently smooth in $x$ and
piece-wise smooth in $t$, and $U_{0}(x)$ is the identity map. Thus,
each point (particle) $x\in\Lambda_{0}$ has its own trajectory in
$R^{d}$: $y(t,x)=U_{t}(x)$, where $y(0,x)=x$ is the initial coordinate
of this particle. It follows from the definition, that the particles
never collide, that is $y(t,x)\neq y(t,x')$ for any $t$ and $x\neq x'$.

$\mathbf{M}_{T}$ is called a system without interaction, if $y(t,x)$
are the solutions of the following equations 
\begin{equation}
\frac{d^{2}y(t,x)}{dt^{2}}=F_{x}(y(t,x)),\ \quad y(0,x)=x,\ \frac{dy(0,x)}{dt}=v(x)\label{mainEqNewton}
\end{equation}
for some given functions: initial velocity $v(x)$ and external forces
$F_{x}(y)$, possibly different for different particles. Further on
we assume that either $F_{x}(y)=F(y)$ does not depend on $x$ or
$F_{x}(y)=\frac{F(y)}{m(x)}$ for some functions $F(y)$ and $m(x)>0$,
see section 4 below. It is always assumed that $v(x)$ and $m(x)$
are sufficiently smooth in $x\in\Lambda_{0}$, and $F(y)$ is smooth
or piece-wise smooth in $y$. Moreover, it is always assumed that
any equation (\ref{mainEqNewton}) has a unique solution on all considered
interval $[0,T)$. Unless otherwise stated, we put $m(x)=1$.

Obviously, the conception of continuum media as consisting of the
continuum number of particles of infinitely small mass, is well known
in mathematics, see for example \cite{Rashevskij}, p. 56. The goal
of this paper is to stress the importance of the notion of regularity,
and give examples of such systems. If the smoothness of $y(t,x)$
follows from general theorems of the theory of ordinary differential
equations, then the main difficulty is to prove the absence of collisions.
We remind that we consider the trajectories not in the phase space
$R^{d}\times R^{d}$, but their projections on the coordinate space
$R^{d}$. The term <<regular>> hints that more general definitions
are possible.

\section{Main results}

\subsection{One-dimensional systems}

\paragraph{Smooth force }

Firstly, note that if $v(x)$ and $F(y)$ are positive and non-decreasing
functions then there will be no collisions, because a particl\d{e}
cannot catch up the particles, which are initially (at $t=0$) to
the right of it.

Now we put $\Lambda_{0}=[0,1]$, and $F(y),y\in[-\infty,\infty),$
is assumed to be smooth. Define the potential energy at any point
$y$ and the full energy of the particle at time $t$, initially at
point $x$, correspondingly to the equation (\ref{mainEqNewton}),
\[
U(y)=-\int_{0}^{y}F(z)dz,H_{t}(x)=\frac{u^{2}(t,y(t,x)))}{2}+U(y(t,x))
\]
where 
\[
u(t,y(t,x))=\frac{dy(t,x)}{dt}=v(t,x)
\]
is the velocity of the particle being at time $t$ at the point $y$.

We shall prove first the simpler, but more intuitive statement, and
later we shall discuss technically more difficult one. Let $T(x,y)$
the first time moment when the point $x\in[0,1]$ will be at the point
$y$.

Assume the following:

1) $v(x)>0$,

2) for all $x\in[0,1]$ and all $y\geq x$ the functions $H_{0}(x)-U(y)>0$.
In particular, this is the case when $F(y)$ is positive for all $y\geqslant0$
(in this case all particles move to the right).

\begin{theorem}\label{Th_dim1_1} Under these assumptions the following
conditions are equivalent: 

1) on the interval $[0,\infty)$ there will not be collisions of particles;

2) for all $y>0$ the functions $T(x,y)$ is strictly decreasing in
$x$ on the interval $x\in[0,\min\{y,1\}]$:

3) for any $y>0$ and for $x\in[0,\min\{y,1\}]$ 
\[
\frac{1}{v(x)}+\frac{v(x)v_{x}'(x)-F(x)}{2\sqrt{2}}\int_{x}^{y}\frac{dz}{((H_{0}(x)-U(z))^{\frac{3}{2}}}\geq0
\]
where the equality is possible only on the discrete subset of points.

\end{theorem}

\begin{theorem}\label{Th_dim1_2} Let now $v(x)\geq0,m(x),F(y)>0\in C^{1}(\mathbb{R}^{1})$.
Then there will not be collisions iff for all $x\in[0,\min\{y,1\}]$
and $y>x$ the following inequality holds: 
\begin{equation}
H'_{0}(x)\left(\frac{1}{\sqrt{2(H_{0}(x)-U(y))}}\frac{1}{F(y)}+\int_{x}^{y}\frac{1}{\sqrt{2(H_{0}(x)-U(z))}}\frac{F'(z)}{F^{2}(z)}dz\right)<\label{nonIntersectCond}
\end{equation}
\[
<\frac{v'(x)\sqrt{m(x)}}{F(x)}+\frac{v(x)m'(x)}{2F(x)\sqrt{m(x)}}
\]
\end{theorem}

In particular, when $v(x)=0,\ m(x)=1$ for all $x\in[0,1]$, the inequality
(\ref{nonIntersectCond}) is equivalent to the following: 
\[
\frac{1}{\sqrt{U(x)-U(y)}}+F(y)\int_{x}^{y}\frac{1}{\sqrt{U(x)-U(z)}}\frac{F'(z)}{F^{2}(z)}dz>0
\]

Note that the similar assertion holds if the functions $F(y),v(x),m(x)$
are piece-wise smooth.

\paragraph{Piece-wise constant force}

\begin{theorem}\label{Th_dim1_constant}\par 1) (one gap) Let for
some $F_{1}>0,F_{2}\geq0$ and $A>1$\par 
\[
F(x)=F_{1},0\leq x<A,\,\,\,F(x)=F_{2},x\geq A
\]
If $v(x)=0$ for all $x\in[0,1]$, then there are no collisions iff
$F_{2}\geq F_{1}$.\par  If $v(x)\geq0$ for all $x\in[0,1]$, then
there will not be collisions iff for all $x\in[0,1]$ both of following
inequalities hold: 
\begin{align}
 & -2(A-x)v'(x)<v(x)+\sqrt{D(x)},\label{stepFirst}\\
 & v'(x)((F_{1}-F_{2})v(x)+F_{2}\sqrt{D(x)})\geqslant F_{1}(F_{1}-F_{2}),\label{stepSecond}
\end{align}
where 
\[
D(x)=v^{2}(x)+2F_{1}(A-x)
\]
\par 2) (two gaps) assume that $v(x)=0$ for all $x\in[0,1]$, and
also that for some $0<F_{2}<F_{1},F_{2}<F_{3}$ and $1<A<B$ 
\[
F(x)=F_{1},0\leq x<A,\,\,\,F(x)=F_{2},x\in[A,B),\,\,\,F(x)=F_{3},x\geq B
\]
Then there will not be collisions iff the following inequality holds:
\begin{equation}
B-A\leqslant\alpha(A-1),\quad\alpha=\frac{F_{1}(F_{3}-F_{1})\left(F_{3}(F_{1}-F_{2})+F_{1}(F_{3}-F_{2})\right)}{(F_{1}-F_{2})^{2}F_{3}^{2}},\label{LtwoStepEqTh}
\end{equation}
\end{theorem}

From this statement the following necessary condition for the absence
of collisions follows: $F_{3}>F_{1}$. Notice also that the set of
all $B>A>1$, for which there will not be collisions, is not empty
under the condition that $F_{3}>F_{1}$.

Unexpected corollary of the second statement of point 1) of the theorem
\ref{Th_dim1_constant} for the case $F_{2}=0$ is the following simple
sufficient (but not necessary) condition for the absence of collisions:
\[
v'(x)\geqslant\frac{F_{1}}{\sqrt{F_{1}x+v^{2}(0)}}.
\]

\subsection{Multi-dimensional systems}

\paragraph{Multi-dimensional analog of monotonicity of the force}

Remind that if the external force does not decrease and initial velocities
also do not decrease then there will not be collisions. We will prove
the following multi-dimensional generalization of this fact.

\begin{theorem} \label{Th_manyDimMonForce} Assume that the force
$F(y)$ is such that for all $x,y\in\mathbb{R}^{d}$ the following
inequality holds: 
\[
(F(y)-F(x),y-x)\geqslant0.
\]
Assume also that for all $x_{1},x_{2}\in\Lambda$ 
\[
(v(x_{2})-v(x_{1}),x_{2}-x_{1})\geqslant0.
\]
Then there will not be collisions on the time interval $[0,\infty)$.
\end{theorem}

\subsubsection*{Linear force}

Assume that the force $F$ is linear, that is 
\[
F(y)=Ay+b,
\]
for some $(d\times d)$-matrix $A$ \textcyr{\char232} $b\in\mathbb{R}^{d}$.

Further on we will assume that all eigenvalues $\lambda_{1},\ldots,\lambda_{d}$
of $A$ are real, and there exists basis $u_{i}$ of the space $\mathbb{R}^{d}$,
consisting of the eigenvectors of $A$, so that $Au_{i}=\lambda_{i}u_{i},\ i=1,\ldots,d$.

\begin{theorem}\label{Th_linear}\par Assume that all eigenvalues
of the matrix $A$ are non negative, and that for all $x_{1},x_{2}\in\Lambda$
\[
(v(x_{2})-v(x_{1}),x_{2}-x_{1})\geqslant0,
\]
Then there will not be collisions.\par \end{theorem}

\subsubsection*{Piece-wise constant force}

\begin{lemma}\label{Lemma_constForce} Let $F(y)=F$ for all $y\in\mathbb{R}^{d}$
and for some constant vector $F\in\mathbb{R}^{d}$. The particle $x_{1},x_{2}$
collide iff the vectors $R(x_{1},x_{2})=x_{2}-x_{1}$ and $V(x_{1},x_{2})=v(x_{2})-v(x_{1})$
are parallel and the following inequality holds: 
\[
(R(x_{1},x_{2}),V(x_{1},x_{2}))<0,
\]
where $(,)$ is the standard euclidean product in $\mathbb{R}^{d}$.\par \end{lemma}

Assume that the force $F$ is defined as: 
\[
F(y)=\begin{cases}
F_{1},\  & y\in\Pi_{1}=\{y=(y^{1},\ldots,y^{d})\in\mathbb{R}^{d}:\ y^{d}<A\}\\
F_{2},\  & y\in\Pi_{2}=\{y=(y^{1},\ldots,y^{d})\in\mathbb{R}^{d}:\ y^{d}\geqslant A\}
\end{cases},
\]
where $F_{k}=(F_{k}^{1},F_{k}^{2},\ldots,F_{k}^{d})\in\mathbb{R}^{d},\ k=1,2$
are constant vectors, and the parameter $A>0$. We shall assume also
that $F_{1}^{d}>0$ for definiteness. Assume also that $\Lambda\subset\Pi_{1}$.

The following statement is both natural and somewhat unexpected generalization
of (\ref{Th_dim1_constant}).

\begin{theorem} \label{Th_stepManyDim} Assume that $v(x)=0$ for
all $x\in\Lambda$ and $F_{2}^{d}\geqslant0$. Then there is no collisions
iff $F_{1}^{d}\leqslant F_{2}^{d}$.\par \end{theorem}

Note that the condition $F_{2}^{d}\geqslant0$ is necessary for the
particle could not return to the set $\Pi_{1}$ after hitting the
set $\Pi_{2}$. Without this condition particle could oscillate between
the sets $\Pi_{1},\Pi_{2}$. Then the analysis becomes more complicated.

\subsubsection*{Central field on the plane}

Let $d=2$ and, besides euclidean $x=(x^{1},x^{2})$ coordinate we
shall use also polar coordinates $(r,\phi)$ on the plane: 
\[
x^{1}=r\cos\phi,\quad x^{2}=r\sin\phi.
\]
Let $\Lambda_{0}$ be bounded and does not contain the origin. Then
it is contained in some annulus 
\[
O(R_{1},R_{2})=\{x:0<R_{1}<r<R_{2}<\infty\}
\]
The force is assumed to be central, that is directed along the radius
vector $\mathbf{r}$ of the point $x$, and equal to 
\[
F(x)=-\frac{\partial U(r)}{\partial r}\frac{\mathbf{r}}{r},\ y\in\mathbb{R}^{2},
\]
where the potential energy $U$ is a smooth scalar function on $(0,\infty)$,
$|\cdot|$ is the euclidean norm.

Denote $r(t,x),\ \phi(t,x)$ the norm and the angle of the point $y(t,x)$
at time moment $t$. Note that the trajectory 
\[
y(t,x)=(r(t,x),\phi(t,x)).
\]
is uniquely defined by the initial velocities field $v(0,x),\ x\in\Lambda$,
or the functions 
\[
\frac{dr(0,x)}{dt},\ \frac{d\phi(0,x)}{dt},\ x\in\Lambda
\]
We need also the following assumptions:
\begin{enumerate}
\item For all points $x\in\Lambda$, the functions 
\[
\frac{dr(0,x)}{dt}=g(|x|)>0.\,\,\,\frac{d\phi(0,x)}{dt}=h(|x|)
\]
depend only on $r$, and the first one is positive. 
\item For all $r_{2}\geqslant r_{1}>R_{1}$ 
\[
-\frac{dU(r_{2})}{dr_{2}}+\frac{M^{2}(r_{1})}{r_{2}^{3}}\geqslant0,
\]
where $M(r)=r^{2}h(r)$ is the kinetic momentum. 
\end{enumerate}
As we will see later, these conditions garanty that $r(t,x)$ monotonically
increases to infinity with $t$.

\begin{theorem}\label{Th_central} Under the formulated assumptions,
for the absence of collisions it is sufficient that for any $R_{1}<r_{1}<R_{2}$
\textcyr{\char232} $r_{2}>r_{1}$ 
\[
\int_{r_{1}}^{r_{2}}\frac{d}{dr_{1}}\frac{1}{\sqrt{2(E_{0}(r_{1})-V(z,r_{1}))}}dz<\frac{1}{g(r_{1})},
\]
where 
\[
E_{0}(r)=\frac{1}{2}g^{2}(r)+U(r)+\frac{1}{2}r^{2}h^{2}(r),\quad V(z,r)=U(z)+\frac{r^{4}h^{2}(r)}{2z^{2}}.
\]
\end{theorem}

Note that the dynamics of $\Lambda_{0}$ can be described as follows.
All intersection points of $\Lambda_{0}$ with the circle $\gamma_{r}$
of radius $r>0$ will become at time $t$ on the circle of some radius
$R(t,r)$, and simultaneously will be rotated around the origin on
the same angle $\phi(t,r)$. Moreover, $R(t,r)$ and $\phi(t,r)$
depend only of $r$ and $t$.

One can do the condition 2) weaker, assuming that $g(|x|)\geqslant0$.
Then the proof should be changed along the plan similar to that in
the theorem \ref{Th_dim1_2}.

\section{Proofs}

\subsection{One dimensional systems}

\paragraph{Smooth force - Theorem \ref{Th_dim1_1}}

The equivalence of 1) and 2) is obvious - this means that no particle
will catch up another particle, situated at time $t=0$ to the right
of it. To prove 3) note that from the energy conservation $H_{t}(x)=H_{0}(x)$
the following formula follows

\begin{equation}
T(x,y)=\int_{x}^{y}\frac{dz}{\sqrt{2(H_{0}(x)-U(z))}}\label{T}
\end{equation}
Note that under our conditions the function $U$ is non-increasing
in $x$, thus the expressions under square root in (\ref{T}) are
always non negative. One has now only to calculate the derivative

\[
\frac{dT(x,y)}{dx}=-\frac{1}{\sqrt{2(H_{0}(x)-U(x))}}+\frac{1}{\sqrt{2}}\int_{x}^{y}\frac{d}{dx}(\frac{1}{((H_{0}(x)-U(z))^{\frac{1}{2}}})dz=
\]
\begin{equation}
=-\frac{1}{v(x)}-\frac{1}{2\sqrt{2}}\int_{x}^{y}\frac{(vv_{x}'-F(x))dz}{((H_{0}(x)-U(z))^{\frac{3}{2}}}\leq0\label{T_v_positive}
\end{equation}

\paragraph{Smooth force - Theorem\ref{Th_dim1_2}}

One can integrate by parts in the formula for $T(x,y)$: 
\[
\sqrt{2}T(x,y)=-\int_{x}^{y}\frac{2}{U'(z)}\ d\sqrt{H_{0}(x)-U(z)}=
\]

\[
=2\left(-\frac{\sqrt{H_{0}(x)-U(y)}}{U'(y)}+\frac{\sqrt{H_{0}(x)-U(x)}}{U'(x)}\right)+2\int_{x}^{y}\sqrt{H_{0}(x)-U(z)}\left(\frac{1}{U'(z)}\right)'dz=
\]

\[
=2h(x,y)+2g(x,y),
\]
where 
\[
h(x,y)=-\frac{\sqrt{H_{0}(x)-U(y)}}{U'(y)}+\frac{\sqrt{H_{0}(x)-U(x)}}{U'(x)}=\frac{\sqrt{H_{0}(x)-U(y)}}{F(y)}-\frac{v(x)\sqrt{m(x)}}{\sqrt{2}F(x)}
\]

\[
g(x,y)=\int_{x}^{y}\sqrt{H_{0}(x)-U(z)}\left(\frac{1}{U'(z)}\right)'dz=\int_{x}^{y}\sqrt{H_{0}(x)-U(z)}\frac{F'(z)}{F^{2}(z)}dz
\]
The derivative of the first summand is: 
\[
\frac{d}{dx}h(x,y)=H'_{0}(x)\frac{1}{2\sqrt{H_{0}(x)-U(y)}}\frac{1}{F(y)}-\frac{v'(x)\sqrt{m(x)}}{\sqrt{2}F(x)}+\frac{v(x)F'(x)\sqrt{m(x)}}{\sqrt{2}F^{2}(x)}-\frac{v(x)m'(x)}{2\sqrt{2}F(x)\sqrt{m(x)}}
\]
Using the known formula: 
\[
\frac{d}{dx}\int_{x}^{y}f(x,z)dz=-f(x,x)+\int_{x}^{y}\frac{\partial}{\partial x}f(x,z)dz.
\]
we get 
\[
\frac{d}{dx}g(x,y)=-\sqrt{H_{0}(x)-U(x)}\frac{F'(x)}{F^{2}(x)}+\frac{1}{2}H_{0}'(x)\int_{x}^{y}\frac{1}{\sqrt{H_{0}(x)-U(z)}}\frac{F'(z)}{F^{2}(z)}dz.
\]
This gives the proof 
\[
\frac{d}{dx}T(x,y)=\sqrt{2}\frac{d}{dx}h(x,y)+\sqrt{2}\frac{d}{dx}g(x,y)=
\]

\[
=H'_{0}(x)\frac{1}{\sqrt{2(H_{0}(x)-U(y))}}\frac{1}{F(y)}-\frac{v'(x)\sqrt{m(x)}}{F(x)}-\frac{v(x)m'(x)}{2F(x)\sqrt{m(x)}}+
\]

\[
+H_{0}'(x)\int_{x}^{y}\frac{1}{\sqrt{2(H_{0}(x)-U(z))}}\frac{F'(z)}{F^{2}(z)}dz
\]

\paragraph{Piece-wise constant force - Theorem \ref{Th_dim1_constant}}

This theorem can be proved using the Theorem \ref{Th_dim1_2} (more
exactly, on its analog for piece-wise smooth force $F(x)$). But the
following simpler proof is more useful.

Let us prove the first assertion of the theorem. Obviously, on the
time interval $[0,\infty)$ there will not be collisions iff $v(T(0,A),x)$
is non decreasing in $x\in[0,1]$. Evidently 
\[
v(T(0,A),x)=v(T(x,A),x)+F_{2}(T(0,A)-T(x,A)),\ \quad v(T(x,A),x)=F_{1}T(x,A).
\]
\[
\frac{dv(T(0,A),x)}{dx}=F_{1}\frac{dT(x,A)}{dx}-F_{2}\frac{dT(x,A)}{dx}=(F_{1}-F_{2})\frac{dT(x,A)}{dx}.
\]
It is clear that $\frac{dT(x,A)}{dx}<0$ for all $x\in[0,1]$, that
gives the assertion.

Let us now prove the second statement. Obviously, on the interval
$[0,\infty)$ there will not be collisions of $v(T(0,A),x)$ is non
decreasing in $x\in[0,1]$ and $T(x,A)$ is decreasing in $x$. Evidently
\[
v(T(0,A),x)=v(T(x,A),x)+F_{2}(T(0,A)-T(x,A)),\ \quad v(T(x,A),x)=v(x)+F_{1}T(x,A).
\]
\begin{equation}
\frac{dv(T(0,A),x)}{dx}=v'(x)+F_{1}\frac{dT(x,A)}{dx}-F_{2}\frac{dT(x,A)}{dx}=v'(x)+(F_{1}-F_{2})\frac{dT(x,A)}{dx}.\label{derivVelocitystep}
\end{equation}
Let us see when the function $T(x,A)$ is decreasing in $x$. From
the equation 
\[
x+v(x)t+\frac{F_{1}}{2}t^{2}=A
\]
we get 
\begin{equation}
T(x,A)=\frac{-v(x)+\sqrt{D(x)}}{F_{1}},\ D(x)=v^{2}(x)+2F_{1}(A-x).\label{D_x}
\end{equation}
Then 
\begin{equation}
\frac{dT(x,A)}{dx}=\frac{-v'(x)+\frac{v(x)v'(x)-F_{1}}{\sqrt{D(x)}}}{F_{1}}=v'(x)\frac{v(x)-\sqrt{D(x)}}{F_{1}\sqrt{D(x)}}-\frac{1}{\sqrt{D(x)}}.\label{dT_dx}
\end{equation}
That is why the condition $\frac{dT(x,A)}{dx}<0$ is equivalent to
the following inequality: 
\[
(v(x)-\sqrt{D(x)})v'(x)<F_{1}.
\]
Multiplying this inequality on $v(x)+\sqrt{D(x)}$, from (\ref{D_x})
we get equivalent inequality: 
\[
-2(A-x)v'(x)<v(x)+\sqrt{D(x)}.
\]
Substituting the formula (\ref{dT_dx}) for $\frac{dT(x,A)}{dx}$
into the formula (\ref{derivVelocitystep}), we get: 
\[
\frac{dv(T(0,A),x)}{dx}=v'(x)\left(1+(F_{1}-F_{2})\frac{v(x)-\sqrt{D(x)}}{F_{1}\sqrt{D(x)}}\right)-\frac{F_{1}-F_{2}}{\sqrt{D(x)}}.
\]
That is why the condition $\frac{dv(T(0,A),x)}{dx}\geqslant0$ is
equivalent to the inequality: 
\[
v'(x)((F_{1}-F_{2})v(x)+F_{2}\sqrt{D(x)})\geqslant F_{1}(F_{1}-F_{2}).
\]
The statement is thus proved.

Let us prove now the third statement. But before the formal proof,
we would like to intuitively explain why the situation is possible
when two infinitely close particles $x_{1}=x,x_{2}=x+dx$ will not
collide after the trajectory of the left point $x_{1}$ will reach
$A$. Assuming that $F_{2}<F_{1}$, the distance between the points
will decrease linearly in time after the moment $T_{1}=T(x_{1},A)$
with velocity $w=v(T_{1},x_{1})-v(T_{1},x_{2})$. As the points are
infinitely close, then $w=a\ dx$ for some constant $a=a(x_{1},x_{2})>0$.
From the other side, the distance $D=y(T_{1},x_{2})-y(T_{1},x_{1})=b\ dx$
between points at time $T_{1}$, for some constant $b=b(x_{1},x_{2})>0$.
That is why the time necessary for the left particle to catch up the
right one, equals $t=t(x_{1},x_{2})=D/w=\frac{b}{a}$. Thus, this
time is already not infinitely small. It ap\d{p}ears that this time
is separated from zero by some constant $t^{*}$ for all $0\leqslant x_{1}<x_{2}\leqslant1$,
that is why it is sufficient to choose the length of the interval
$B-A$, where the force $F_{2}$ acts, so that any point from $[0,1]$
pass the interval $[A,B]$ for the time less than $t^{*}$. It remains
to choose the force $F_{3}$ sufficiently large.

Let $T=T(0,B)$. It is clear that on the interval $[0,+\infty]$ there
will not be collisions iff $v(T,x)$ is non decreasing in $x\in[0,1]$.
We have evident equalities: 
\[
v(T,x)=v(T(x,B),x)+F_{3}(T-T(x,B)),
\]

\[
v(T(x,B),x)=v(T(x,A),x)+F_{2}(T(x,B)-T(x,A)),\quad v(T(x,A),x)=F_{1}T(x,A).
\]
This gives 
\[
v(T,x)=F_{1}T(x,A)+F_{2}(T(x,B)-T(x,A))+F_{3}(T-T(x,B)).
\]
\[
\frac{dv(T,x)}{dx}=F_{1}\frac{dT(x,A)}{dx}+F_{2}(\frac{dT(x,B)}{dx}-\frac{dT(x,A)}{dx})-F_{3}\frac{dT(x,B)}{dx}=
\]
\[
=(F_{1}-F_{2})\frac{dT(x,A)}{dx}-(F_{3}-F_{2})\frac{dT(x,B)}{dx}.
\]
We can find $T(x,B)$. Clearly 
\[
y(T(x,A)+s,x)=A+T(x,A)F_{1}s+\frac{s^{2}}{2}F_{2},
\]
for $0\leqslant s\leqslant T(x,B)-T(x,A)$. Then from the condition
$y(x,T(x,A)+s)=B$ we find that 
\[
s=s(x)=\frac{-T(x,A)F_{1}+\sqrt{D(x)}}{F_{2}},\ D(x)=T^{2}(x,A)F_{1}^{2}+2(B-A)F_{2}.
\]
This gives 
\[
T(x,B)=T(x,A)+s(x)=\frac{-T(x,A)(F_{1}-F_{2})+\sqrt{D(x)}}{F_{2}}.
\]
Let us find the derivative: 
\[
\frac{dT(x,B)}{dx}=\frac{-\frac{dT(x,A)}{dx}(F_{1}-F_{2})+\frac{T(x,A)\frac{dT(x,A)}{dx}F_{1}^{2}}{\sqrt{D(x)}}}{F_{2}}=-\frac{dT(x,A)}{dx}\frac{(F_{1}-F_{2})-\frac{T(x,A)F_{1}^{2}}{\sqrt{D(x)}}}{F_{2}}.
\]
Then 
\[
\frac{dv(T,x)}{dx}=\frac{dT(x,A)}{dx}\left((F_{1}-F_{2})+(F_{3}-F_{2})\frac{(F_{1}-F_{2})-\frac{T(x,A)F_{1}^{2}}{\sqrt{D(x)}}}{F_{2}}\right)=
\]
\[
=\frac{dT(x,A)}{dx}\left(\frac{(F_{1}-F_{2})F_{3}}{F_{2}}-(F_{3}-F_{2})\frac{T(x,A)F_{1}^{2}}{F_{2}\sqrt{D(x)}}\right)
\]
As $\frac{dT(x,A)}{dx}<0$, then the condition $\frac{dv(T,x)}{dx}\geqslant0$
is equivalent to the inequality: 
\[
T(x,A)\geqslant\beta\sqrt{D(x)},\ \beta=\frac{(F_{1}-F_{2})F_{3}}{(F_{3}-F_{2})F_{1}^{2}}.
\]
Taking the square of the latter inequality, after some transformations
\[
T^{2}(x,A)\geqslant\beta^{2}(T^{2}(x,A)F_{1}^{2}+2(B-A)F_{2}).
\]
we get the equivalent inequality

\begin{equation}
T^{2}(x,A)\geqslant\frac{2(B-A)F_{2}\beta^{2}}{1-\beta^{2}F_{1}^{2}}.\label{L201604211}
\end{equation}
This condition should hold for all $x\in[0,1]$. But $T(x,A)\geqslant T(1,A)$
for all $x\in[0,1]$. Hence, (\ref{L201604211}) is equivalent to
the inequality: 
\[
T^{2}(1,A)\geqslant\frac{2(B-A)F_{2}\beta^{2}}{1-\beta^{2}F_{1}^{2}}.
\]
Substituting the expression (\ref{D_x}) for $T(1,A)$, with $v(x)=0,x=1$,
to this inequality, we get 
\[
(A-1)\frac{(1-\beta^{2}F_{1}^{2})}{F_{1}F_{2}\beta^{2}}\geqslant B-A.
\]
Transforming the second factor in the left side of this inequality,
we get: 
\[
\alpha=\frac{(1-\beta^{2}F_{1}^{2})}{F_{1}F_{2}\beta^{2}}=\frac{1}{F_{1}F_{2}}\left(\frac{1}{\beta}-F_{1}\right)\left(\frac{1}{\beta}+F_{1}\right)=
\]

\[
=\frac{F_{1}^{2}}{F_{1}F_{2}(F_{1}-F_{2})^{2}F_{3}^{2}}\left((F_{3}-F_{2})F_{1}-(F_{1}-F_{2})F_{3}\right)\left((F_{3}-F_{2})F_{1}+(F_{1}-F_{2})F_{3}\right)=
\]

\[
=\frac{F_{1}}{(F_{1}-F_{2})^{2}F_{3}^{2}}\left(F_{3}-F_{1}\right)\left((F_{3}-F_{2})F_{1}+(F_{1}-F_{2})F_{3}\right)
\]
The proof is finished.

\subsection{Mufti-dimensional systems}

\paragraph{Proof of Theorem \ref{Th_manyDimMonForce}}

For any two unequal points $x_{1},x_{2}\in\Lambda$ consider the function:
\[
r(t)=||y(t,x_{2})-y(t,x_{1})||^{2}=(y(t,x_{2})-y(t,x_{1}),y(t,x_{2})-y(t,x_{1})).
\]
Differentiation gives 
\[
\frac{d^{2}r(t)}{dt^{2}}=2(F(y(t,x_{2}))-F(y(t,x_{1})),y(t,x_{2})-y(t,x_{1}))+2||v(t,x_{2})-v(t,x_{1})||^{2}.
\]
Using the conditions on $F$ we get 
\[
\frac{d^{2}r(t)}{dt^{2}}\geqslant0.
\]
For initial conditions 
\[
r(0)=||x_{2}-x_{1}||^{2}>0,\frac{dr}{dt}(0)=2(v(x_{2})-v(x_{1}),x_{2}-x_{1})\geqslant0.
\]
Our statement follows from these three inequalities .

\subsubsection*{Linear force - proof of Theorem \ref{Th_linear}}

We will show that for any $x$ the quadratic form 
\begin{equation}
Q(x)=(Ax,x)\geq0\label{Q_positive}
\end{equation}
As any $x\in\mathbb{R}^{d}$ can be uniquely written as 
\[
x=\sum_{i=1}^{d}x_{i}u_{i},\ x_{i}\in\mathbb{R}.
\]
then we can define the symmetric matrix $S=(s_{i,j})$ by 
\[
s_{i,j}=\frac{1}{2}(\lambda_{i}(u_{i},u_{j})+\lambda_{j}(u_{j},u_{i}))=\frac{1}{2}(\lambda_{i}+\lambda_{j})(u_{i},u_{j}).
\]
As 
\[
Q(x)=\sum_{i,j}\lambda_{i}(u_{i},u_{j})x_{i}x_{j}=(Sx,x),
\]
we can write 
\[
S=\Lambda U+U\Lambda,
\]
where $\Lambda=\mathrm{diag}(\lambda_{1},\ldots,\lambda_{d})$ is
the diagonal matrix and $U=((u_{i},u_{j}))$. That is why the matrix
$S$ is non negative definite, whence (\ref{Q_positive}) follows.
Now, using Theorem \ref{Th_manyDimMonForce} we get the final statement.

\paragraph{Piece wise constant force. Proof of Lemma \ref{Lemma_constForce}}

We have the evident equality for the constant force 
\[
y(t,x)=x+v(x)t+\frac{Ft^{2}}{2}.
\]
Then 
\[
y(t,x_{2})-y(t,x_{1})=V(x_{1},x_{2})t+R(x_{1},x_{2}).
\]
The statement follows.

\paragraph{Proof of Theorem \ref{Th_stepManyDim}}

From theorem \ref{Th_dim1_1} for one-dimensional case it follows
that $F_{1}^{d}>F_{2}^{d}$ is a necessary for existence of collisions.
For $x=(x^{1},\ldots,x^{d})\in\Lambda$ denote 
\[
T(x)=\sqrt{\frac{2(A-x^{d})}{F_{1}^{d}}},
\]
the time moment when $(y(t,x),e_{d})=A$.

Consider two points $x_{1}=(x_{1}^{1},\ldots,x_{1}^{d})\in\Lambda,\ x_{2}=(x_{2}^{1},\ldots,x_{2}^{d})\in\Lambda$.
Let $x_{2}^{d}>x_{1}^{d}$. By our assumptions we have $T(x_{1})>T(x_{2})$.
Besides that it is clear before time moment $T(x_{1})$ the points
$x_{1},x_{2}$ will not collide. Starting from the moment $T(x_{1})$
we are in the situation of Lemma \ref{Lemma_constForce}. In fact,
we have at time moment $T(x_{1})$: 

\selectlanguage{russian}%
\inputencoding{koi8-r}\[
y(T(x_{1}),x_{1})=x_{1}+F_{1}\frac{T^{2}(x_{1})}{2},\quad v(T(x_{1}),x_{1})=F_{1}T(x_{1}).
\]
\[
y(T(x_{1}),x_{2})=y(T(x_{2}),x_{2})+v(T(x_{2}),x_{2})s+F_{2}\frac{s^{2}}{2},\quad v(T(x_{1}),x_{2})=v(T(x_{2}),x_{2})+F_{2}s,\ 
\]
where 
\[
s=T(x_{1})-T(x_{2}),\ v(T(x_{2}),x_{2})=F_{1}T(x_{2}),\ y(T(x_{2}),x_{2})=x_{2}+F_{1}\frac{T^{2}(x_{2})}{2}.
\]
Then 
\[
V(x_{1},x_{2})=v(T(x_{1}),x_{2})-v(T(x_{1}),x_{1})=F_{1}(T(x_{2})-T(x_{1}))+F_{2}s=s(F_{2}-F_{1}).
\]
\[
R(x_{1},x_{2})=y(T(x_{1}),x_{2})-y(T(x_{1}),x_{1})=x_{2}+F_{1}\frac{T^{2}(x_{2})}{2}+F_{1}T(x_{2})s+F_{2}\frac{s^{2}}{2}-x_{1}-F_{1}\frac{T^{2}(x_{1})}{2}=
\]
\[
=(x_{2}-x_{1})+F_{1}\left(\frac{T^{2}(x_{2})}{2}-\frac{T^{2}(x_{1})}{2}+T(x_{2})s\right)+F_{2}\frac{s^{2}}{2}=(x_{2}-x_{1})-F_{1}\frac{s^{2}}{2}+F_{2}\frac{s^{2}}{2}=(x_{2}-x_{1})+\frac{s}{2}V(x_{1},x_{2}).
\]

\selectlanguage{english}%
\inputencoding{latin9}Thus we get 
\[
R(x_{1},x_{2})=x_{2}-x_{1}+\frac{s}{2}V(x_{1},x_{2}).
\]
It follows that the vectors $R,V$ are parallel iff the vector $x_{2}-x_{1}$
is parallel to the vector $F_{2}-F_{1}$. Assume that $x_{1}$ is
an internal point of $\Lambda$. Put $x_{2}=x_{1}+h(F_{2}-F_{1})$,
where of course $h<0$, as $x_{2}^{d}>x_{1}^{d},\ F_{2}^{d}<F_{1}^{d}$.
It is clear that for $|h|$ sufficiently small the point $x_{2}$
will belong to $\Lambda$. We have the equality: 
\[
(R,V)=s\left(h||F_{2}-F_{1}||^{2}+\frac{s^{2}}{2}||F_{2}-F_{1}||^{2}\right)=s||F_{2}-F_{1}||^{2}\left(h+\frac{s^{2}}{2}\right)=s||F_{2}-F_{1}||^{2}\left(h+\bar{\bar{o}}(h)\right)
\]
as $h\rightarrow0-$. The conclusion is that there exists $h<0$ such
that $(R,V)<0$. Thus, the points $x_{1},x_{2}$ should collide. That
gives the proof.

\subsubsection*{Central field on the plane}

\paragraph{Proof of Theorem \ref{Th_central}}

We remind some known facts concerning particle motion in the central
field. In this case the kinetic (angular) momentum 
\[
M(t,x)=M(x)=M(|x|)=r^{2}(t,x)\frac{d\phi(t,x)}{dt},
\]
does not depend on time and equals 
\begin{equation}
M(x)=|x|^{2}h(|x|).\label{kinMomFor}
\end{equation}
Dynamics of the radius vector of $x$ is defined by the equation 
\begin{equation}
\frac{d^{2}r(t,x)}{dt^{2}}=-\frac{\partial E}{\partial r},\ r(0,x)=|x|,\ \frac{dr(0,x)}{dt}=g(|x|),\label{normEq}
\end{equation}
where 
\[
E=\frac{1}{2}\left(\frac{dr(t,x)}{dt}\right)^{2}+V(r(t,x)),
\]
and effective potential energy is defined as 
\[
V(r(t,x))=V(r(t,x),x)=U(r(t,x))+\frac{M^{2}(x)}{2r^{2}(t,x)}=V(r(t,x),|x|).
\]
For any two points $x_{1},x_{2}\in\Lambda$ consider two cases:
\begin{enumerate}
\item $|x_{1}|=|x_{2}|=r$. Then $r(t,x_{1})=r(t,x_{2})$ for any $t\geqslant0$.
In fact, the equality (\ref{kinMomFor}) shows that the functions
$r(t,x_{1})$ \textcyr{\char232} $r(t,x_{2})$ satisfy the same differential
equation (\ref{normEq}) with the same initial conditions. Moreover,
by conservation of kinetic momenta: 
\[
\phi(t,x_{i})=M(r)\int_{0}^{t}\frac{1}{r^{2}(s,x_{i})}ds+\phi(0,x_{i}),\ i=1,2.
\]
It follows that the angles between the points $x_{1},x_{2}$ are conserved,
that implies the absence of collisions. 
\item $|x_{1}|<|x_{2}|$. In this case the proof is similar to the one-dimensional
interval case discussed above. 
\end{enumerate}
By conditions 1) and 2) the norm of $x$ monotonically increases.
Denote $T_{r_{1}}(r_{2})$ the time moment when the particle, moving
in the field of effective potential energy $V(r,|r_{1}|)$ with initial
conditions $r(0)=r_{1},\ \frac{dr(0)}{dt}=g(r_{1})$, reaches point
$r_{2}$. As earlier, we have: 
\[
T_{r_{1}}(r_{2})=\int_{r_{1}}^{r_{2}}\frac{dz}{\sqrt{2\left(E_{0}(r_{1})-V(z,M(r_{1}))\right)}},\quad E_{0}(r)=\frac{1}{2}g^{2}(r)+V(r,r).
\]
It is clear that if $T_{r_{1}}(r_{2})$ is decreasing in $r_{1}$
for any $r_{1}\leqslant r_{2}$ and $R_{1}<r_{1}<R_{2}$, then there
will not be collisions. From this formula for the derivative 
\[
\frac{dT_{r_{1}}(r_{2})}{dr_{1}}=-\frac{1}{g(r_{1})}+\int_{r_{1}}^{r_{2}}\frac{d}{dr_{1}}\frac{1}{\sqrt{2\left(E_{0}(r_{1})-V(z,M(r_{1}))\right)}}\ dz.
\]
the assertion follows.

\section{Conclusion}

Here we do several comments about further problems (a lot of them)
concerning the density, possible interactions in such systems and
Euler equation.

\paragraph{Density}

Consider the case when $F_{x}(y)=F(y)$ does not depend on $x$. The
density at time $t=0$ is defined as arbitrary smooth positive function
$\rho(0,x)$ on $\Lambda_{0}$, and the density at time $t$ on $\Lambda_{t}$
as 
\[
\rho(t,y)=\rho(0,U_{t}^{-1}y)
\]
It is well known (one line proof) that it satisfies the famous conservation
law (Liouville equation)

\begin{equation}
\rho_{t}+(u\rho)_{x}=\rho_{t}+u\rho_{x}+\rho u_{x}=0\label{mass_conservation_1}
\end{equation}

Note that in all our examples the density monotonically tends to zero.
Let us give examples when it tends to infinity.

Consider a smooth curve $z(t),t\in[0,\infty)$ such that: $z(0)=1$,
$z(t)>0$ for any $t\in[0,\infty)$, $z(t)\to0$ as $t\to\infty$
and $z'(t)<0$, that is $z(t)$ is strictly decreasing.

Then consider the system with $\Lambda_{0}=(0,1]$, putting 
\[
v(x)=z'(t),F(y)=z''(t)
\]
where $t$ is uniquely defined from the condition $z(t)=x$. Otherwise
speaking, positive force decreases the velocities to zero. Moreover,
the particles never get the point $x=0$.

\paragraph{About more general regular systems}

Any function $m(x)$ on $\Lambda_{0}$ can be considered as the mass
or charge density, givings some links to real physical forces - gravitational
and electrostatic. More general forces $F_{x}(y)$, different for
different particles, does not seem interesting due to the following
simple theorem.

Assume that the following non-recurrence conditions holds: for any
pair of points $x,z$ the trajectory $y(t,x)$ passes the point $z$
mot more than once.

\begin{proposition}\par Then the regular continuum system can be
presented as the system without interaction with external forces $F_{x}(y)$.\par \end{proposition}

In fact, consider any trajectory $y(t,x)=U_{t}x$ with initial conditions
(\ref{mainEqNewton}). Then it is sufficient to define the force,
acting on the particle $x$, at point $y=y(t,x)$, as 
\[
F_{x}(y)=\frac{d^{2}y(t,x)}{dt^{2}}
\]
so that the system defined by the diffeomorphisms $U_{t}$ were no
interaction system.

Now we say very shortly what could be continuum systems with interaction.
Interaction in such systems can local, when the force acting on the
particle at point $y$ at time moment $t$, looks like 
\begin{equation}
F(y)=f(\rho(y),\nabla\rho(y))\label{local}
\end{equation}
and non-local with the force 
\begin{equation}
F(y)=\int g(|z|,\rho(y),\rho(y+z))dz\label{nonlocal}
\end{equation}
for some functions $f$ and $g$.

Of course, the introduced systems are approximations for the corresponding
finite $N$ particle (with very large $N$) systems. Intuitively one
say that in case (\ref{local}) these systems can be approximations
for $N$-particle systems with two particle interaction decaying at
infinity, that is where only interaction with bounded (not depending
on $N$) number of particles is essential). But the case (\ref{nonlocal})
can be approximation of systems with so called mean field interaction,
where each particle interacts with the number of particles of the
order $N$. Some concrete examples are in progress.

\paragraph{Euler equations and characteristics}

Regularity condition says that for any given $t,y\in\Lambda_{t},$
there is not more than one particle $x\in\Lambda_{0}$ with $y=y(t,x)$,
that is the velocity field is unambiguously defined 
\[
u(t,y(t,x))=\frac{dy(t,x)}{dt}
\]
Then it is easy to prove that this velocity field, for regular system
without interaction, satisfies the Eular equation 
\begin{equation}
\frac{\partial u(t,y)}{\partial t}+\sum_{\alpha}\frac{\partial u(t,y)}{\partial y_{\alpha}}u_{\alpha}(t,y)=F(y)\label{EulerEq}
\end{equation}
In fact, the acceleration of the particle with trajectory $y(t,x)$
equals 
\begin{equation}
\frac{du_{i}(t,y(t,x))}{dt}=\frac{\partial u_{i}(t,y(t,x))}{\partial t}+\sum\frac{\partial u_{i}(t,y(t,x))}{\partial y_{j}}u_{j}(t,y(t,x)).\label{euler_accel}
\end{equation}
thus equal to the force $F(y(t,x))$. We repeat once more that the
absence of collisions allows such derivation of the Euler equation.

Consider now the Euler equation as the abstract partial differential
equation. The Cauchy problem for it with $t\in[0,\infty),y=(y_{1},...,y_{d})\in R^{d},$
and initial conditions 
\[
u(0,x)=v(x)
\]
was studied in many papers. Part of these results one can find in
textbooks and monographs, see \cite{Arnold_1,Arnold_2,Kruzhkov_trudy,Kruzhkov_posobie,Filippov,Rashevskij,Bogoliubov,ChorinMarsden,MarchPulvirenti,Slezkin,Talman,Temam}.

Let now $\Lambda_{0}$ be the real axis $\mathbb{R}$, and consider
the equation (\ref{mainEqNewton}), with similar assumptions on the
functions $F(y),v(x)$ for all $y,x\in\mathbb{R}$.

It is well known that the characteristics $y(t,x)$ are parametrized
by the points $x\in\mathbb{R}$ and satisfy the Newton equation, describing
thus the movement of particles under the external force $F(y)$. Moreover,
the structure of the set of characteristics (more exactly, the projection
of this set on the $x$-space) defines existence and uniqueness for
the Cauchy problem. However, in the general case the conditions for
the absence of collisions are quite non trivial, and now only examples
of this structure can be well understood.

Consider the following example (see \cite{Filippov}): $F(y)=0$ for
all $y\in\mathbb{R}$ and $v(x)=-arctg(x)$. It can be proved that
in this case the solution $u(t,y)$ of equation (\ref{EulerEq}) exists
for any $t\leqslant1,\ y\in\mathbb{R}$, but it cannot be continuously
prolongated to the domain $t>1$. Correspondingly, one can prove that
first collisions in this system appear at time moment $t=1$. This
example of $v(x)$ is a particular case of our remark, concerning
the monotonicity of $v(x)$, in the beginning of the section 2.1. 

The following more general result follows from our theorem \ref{Th_dim1_2}.

\begin{theorem} Let $v(x)\geq0,F(y)\in C^{2}(\mathbb{R})$ and $F(y)>0$
for any $y\in\mathbb{R}$. The equation (\ref{EulerEq}) has a smooth
solution $u(t,y)$ for $t\geqslant0,\ y\in\mathbb{R}$, with initial
condition $u(0,x)=v(x)$, iff for all $x<y$ the following inequality
holds: 
\[
H'_{0}(x)\left(\frac{1}{\sqrt{2(H_{0}(x)-U(y))}}\frac{1}{F(y)}+\int_{x}^{y}\frac{1}{\sqrt{2(H_{0}(x)-U(z))}}\frac{F'(z)}{F^{2}(z)}dz\right)<\frac{v'(x)}{F(x)}
\]
\end{theorem}

One can consider also the Cauchy problem in bounded domains with moving
boundary, consisting of two points $L(t)=y(t,0),\ R(t)=y(t,1)$, and
some boundary conditions on it, which are the Newton equations 
\[
\frac{d^{2}L}{dt^{2}}=F(L),\frac{d^{2}R}{dt^{2}}=F(R)
\]
with initial conditions
\[
L(0)=0,R(0)=1,\frac{dL}{dt}(0)=v(0),\frac{dR}{dt}(0)=v(1).
\]


\begin{thebibliography}{10}
\bibitem{Arnold_1}Arnold V. I. Lectures on partial differential equations.
1995. Moscow.

\bibitem{Arnold_2}Arnold V. I. Dopolnitelnye glavy teorii obyknovennyh
differentsialnyh uravnenij. 1978. Moscow.

\bibitem{Kruzhkov_trudy}Kruzhkov S. N. Selected papers. 2000. Moscow.
Fizmatlit.

\bibitem{Kruzhkov_posobie}Goritskij A. Yu., Kruzhkov S. N., Chechkin
G. A. First order partial differential equations (textbook). 1999.

\bibitem{Filippov}Filippov A. F. Introduction to differential equations.
2007. Moscow.

\bibitem{Bogoliubov}Bogolyubov N. A. About some statistical methods
in mathematical physics. 1945, Kiev\.{ }Academy of Sciences of Ukrainian
SSR.

\bibitem{Rashevskij}Rashevsky P. K. Riemann geometry and tensor analysis.
1967. Moscow.

\bibitem{ChorinMarsden}Chorin A., Marsden J.. A mathematical introduction
to fluid mechanics. Springer. 1992.

\bibitem{MarchPulvirenti}Marchioro C., Pulvirenti M. Mathematical
Theory of Incompressible Nonviscous Fluids. Springer. 1993.

\bibitem{Temam}Temam R., Miranville A. Mathematical modeling in continuum
mechanics. Cambridge Univ. Pfress, 2005.

\bibitem{Talman}Talman R. Geometric mechanics. WILEY, Second ed.
2007.

\bibitem{Slezkin}Slezkin N. A. Lectures on molecular hydrodynamics.
Moscow State University\textcyr{\char200\char231\char228}. 1981. \end{thebibliography}
\end{document}